\documentclass[lettersize,journal]{IEEEtran}
\usepackage{amsmath,amsfonts}
\usepackage{algorithmic}
\usepackage{array}
\usepackage{diagbox}
\usepackage{color}
\usepackage{amssymb}
\usepackage{pifont}
\usepackage{mathabx}
\usepackage{enumitem}
\usepackage{bm}
\makeatletter
\renewcommand\subsubsection{\@startsection{subsubsection}{3}{\z@}%
 {.5ex \@plus .1ex \@minus .05ex}%
 {.1ex \@plus .1ex}%
 {\normalfont\normalsize\itshape}}

\makeatother
\usepackage[caption=false,font=normalsize,labelfont=scriptsize,textfont=scriptsize]{subfig}
\usepackage{textcomp}
\usepackage{stfloats}
\usepackage{makecell}
\usepackage{url}
\usepackage{verbatim}
\usepackage{graphicx}
\usepackage{multirow}
\usepackage{threeparttable} 
\def\BibTeX{{\rm B\kern-.05em{\sc i\kern-.025em b}\kern-.08em
 T\kern-.1667em\lower.7ex\hbox{E}\kern-.125emX}}
\usepackage{balance}
\begin{document}
\title{Stacked Intelligent Metasurfaces for Wireless Communications: Applications and Challenges}
\author{Hao Liu, Jiancheng An, \IEEEmembership{Member, IEEE}, Xing Jia, Lu Gan, \IEEEmembership{Member, IEEE},\\George K. Karagiannidis, \IEEEmembership{Fellow, IEEE}, Bruno Clerckx, \IEEEmembership{Fellow, IEEE}, Mehdi Bennis, \IEEEmembership{Fellow, IEEE},\\M\'{e}rouane Debbah, \IEEEmembership{Fellow, IEEE}, and Tie Jun Cui, \IEEEmembership{Fellow, IEEE}
\thanks{This work is supported by National Natural Science Foundation of China 62471096. (Corresponding Author: Jiancheng An)

H. Liu, X. Jia, and L. Gan are with the School of Information and Communication Engineering, University of Electronic Science and Technology of China (UESTC), Chengdu, Sichuan 611731, China. L. Gan is also with the Yibin Institute of UESTC, Yibin, Sichuan 644000, China (e-mail: liu.hao@std.uestc.edu.cn, xingjia1999@163.com, ganlu@uestc.edu.cn). J. An is with the School of Electrical and Electronics Engineering, Nanyang Technological University, Singapore 639798 (e-mail: jiancheng.an@ntu.edu.sg). G. K. Karagiannidis is with the Department of Electrical and Computer Engineering, Aristotle University of Thessaloniki, 54124 Thessaloniki, Greece (e-mail: geokarag@auth.gr). B. Clerckx is with the Department of Electrical and Electronic Engineering, Imperial College London, London SW7 2AZ, U.K. (e-mail: b.clerckx@imperial.ac.uk). M. Bennis is with the Center for Wireless Communications, Oulu University, Oulu 90014, Finland (e-mail: mehdi.bennis@oulu.fi). M. Debbah is with KU 6G Research Center, Khalifa University of Science and Technology, P O Box 127788, Abu Dhabi, UAE, and CentraleSupelec, University Paris-Saclay, 91192 Gif-sur-Yvette, France (e-mail: merouane.debbah@ku.ac.ae). T. Cui is with the State Key Laboratory of Millimeter Wave, Southeast University, Nanjing 210096, China (e-mail: tjcui@seu.edu.cn).}}
\maketitle
\begin{abstract}
The rapid growth of wireless communications has created a significant demand for high throughput, seamless connectivity, and extremely low latency. To meet these goals, a novel technology -- stacked intelligent metasurfaces (SIMs) -- has been developed to perform signal processing by directly utilizing electromagnetic waves, thus achieving incredibly fast computing speed while reducing hardware requirements. In this article, we provide an overview of SIM technology, including its underlying hardware, benefits, and exciting applications in wireless communications. Specifically, we examine the utilization of SIMs in realizing transmit beamforming and semantic encoding in the wave domain. Additionally, channel estimation in SIM-aided communication systems is discussed. Finally, we highlight potential research opportunities and identify key challenges for deploying SIMs in wireless networks to motivate future research.
\end{abstract}
\begin{IEEEkeywords}
Stacked intelligent metasurfaces (SIMs), wave-domain signal processing, electromagnetic neural network, semantic communication, channel estimation.
\end{IEEEkeywords}
\section{Introduction}
The exponential growth of wireless network scales has triggered a surging need for ultra-high data rates, seamless connectivity, and extremely low latency, driving research efforts toward innovative transceiver architectures and advanced communication techniques to enhance network capacity, energy efficiency, and reliability. Over the decades, multiple-input multiple-output (MIMO) technologies have completely revolutionized wireless communication systems \cite{zhang2023RHS}. By integrating multiple antennas at transceivers, MIMO systems are capable of harnessing multiplexing gain to boost spectrum efficiency, while leveraging spatial diversity to enhance link reliability. Nonetheless, the scalability of MIMO systems comes with increasing challenges in practical deployments, as the utilization of a large number of antennas results in high hardware complexity and elevated energy consumption.

To address these challenges, hybrid MIMO architectures have emerged as a promising solution. By utilizing a small number of radio frequency (RF) chains and cascading with analog phase shifters, hybrid MIMO systems can strike flexible trade-offs between system complexity and performance \cite{Hu_cellfree}. Recently, reconfigurable holographic surfaces (RHSs) have been utilized to further enhance the energy efficiency of hybrid MIMO implementations. Specifically, an RHS consists of a two-dimensional metasurface densely packed with many sub-wavelength radiating meta-atoms that can manipulate electromagnetic waves in a programmable manner \cite{zhang2023RHS}. When integrated into transceivers, the nearly continuous aperture of RHS facilitates the formation of highly directional beams. However, hybrid MIMO systems face inherent hardware constraints in fully achieving the spatial diversity and multiplexing gain offered by large-scale antenna arrays or metasurfaces. Moreover, the single-layer metasurface structure limits the signal processing capabilities of RHSs, thus still requiring digital beamforming to suppress interference among users.

To overcome these limitations, stacked intelligent metasurfaces (SIMs) have emerged as a novel technology for transforming wireless transceiver design \cite{li2021spectrally, wang2024multi, liu2022programmable}. In general, an SIM is composed of multiple layers of programmable metasurfaces, each embedded with many low-cost meta-atoms that are capable of independently modulating the amplitude and phase of incident electromagnetic waves. Thanks to the quasi-neural network architecture, SIM is an advanced analog computing platform capable of performing complex signal processing tasks directly in the wave domain \cite{li2021spectrally, liu2022programmable}. Moreover, programmable metasurfaces allow SIM to dynamically adjust its electromagnetic response in real-time, ensuring adaptability to diverse propagation environments and computational requirements. Recent research efforts have demonstrated the capabilities of SIM in performing direction-of-arrival (DOA) estimation \cite{an2024magazine} and MIMO precoding \cite{liu2024drl}, offering simplified hardware design while significantly reducing latency and cost.

While SIMs hold great promise for executing diverse computing tasks, their practical implementation faces a range of new technological hurdles. Against this background, this article presents a comprehensive overview of SIM technology, focusing on its prospective applications in wireless communication systems. We begin by exploring the potential of SIMs for realizing wave-domain beamforming and semantic encoding. Moreover, channel estimation in SIM-assisted communication systems is discussed. Finally, we identify several promising opportunities and critical research challenges that need to be overcome to fully harness the capabilities of SIMs. 

\section{Functionality, Hardware, and Deployment\\of SIMs}
In contrast to conventional digital signal processors, SIMs have a unique advantage: they can process complex electromagnetic signals directly in the wave domain. This drastically reduces both energy consumption and processing delays required associated with matrix computations \cite{an2024magazine}. When electromagnetic waves pass through these layered metasurfaces, each meta-atom functions as a secondary radiation source, transmitting signals to all the meta-atoms on the next layer \cite{liu2022programmable}. As a consequence, a neural network-like architecture is formed, where each meta-atom on a metasurface layer operates as an artificial neuron, processing and sending information to the subsequent layer. This sophisticated physical architecture empowers SIMs to execute parallel computation tasks at exceptionally high speeds.

As illustrated in Tab. \ref{Hard_survey}, existing SIM prototypes can be classified into three hardware (\textbf{H}) types based on their programmability and integration with active amplifiers. In the following, we delve into the distinctive features of these three hardware types and corresponding application scenarios in detail.

\begin{table}[!t]
\centering
\caption{Illustration of three typical SIM hardware types.}
\includegraphics[width = 8.8 cm]{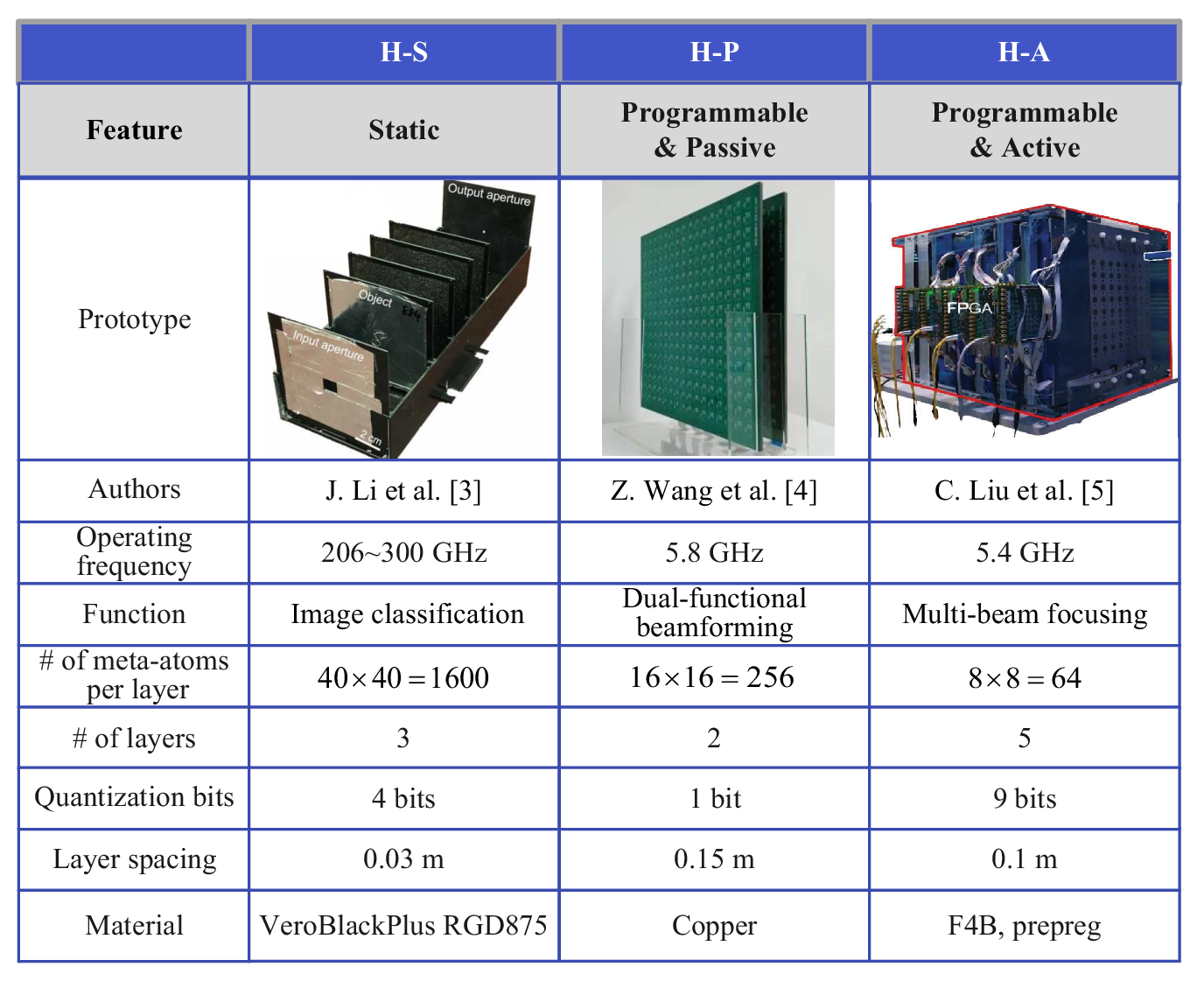}
\label{Hard_survey}
\end{table}

\begin{itemize}
\item \textbf{H-S:} Static (\textbf{S}) SIMs are designed without control circuits, resulting in fixed interconnection structures once manufactured. Static SIMs can manipulate the electromagnetic properties of propagating waves without requiring additional energy for wave-domain computations. As depicted in Tab. \ref{Hard_survey}, a static SIM employs three closely placed diffractive metasurfaces and a single detector operating across multiple frequencies to achieve multi-class image classification \cite{li2021spectrally}. Each layer is fabricated using 3D-printed material -- VeroBlackPlus RGD875. Moreover, the SIM designed in \cite{li2021spectrally} was seamlessly integrated with a shallow electronic neural network to recover images based on the compressed energy distribution patterns. Due to their non-programmable nature, static SIMs are cost-effective and particularly suitable for processing local tasks that are robust against environmental changes, such as DOA estimation in suburban districts \cite{an2024magazine}. However, this architecture needs to be redesigned and remanufactured once conditions or tasks change.
\item\textbf{H-P:} Programmable passive (\textbf{P}) SIMs can be reconfigured using a field programmable gate array (FPGA), which allows them to adapt to environmental fluctuations and handle various computing needs. Once configured, these passive SIMs can perform the desired functions without additional energy consumption. Tab. \ref{Hard_survey} illustrates a programmable passive SIM designed for an integrated sensing and communication (ISAC) system that operates at $5.8$ GHz \cite{wang2024multi}. Specifically, each metasurface consists of a receiving layer, a bias layer, a ground plane, and a radiating layer, with three substrate layers positioned between them. Furthermore, each meta-atom can apply a $1$-bit discrete phase shift to effectively shape electromagnetic wavefronts. Due to their reconfigurability, programmable passive SIMs show significant potential for enabling fully analog beamforming, thus reducing RF power consumption and simplifying MIMO transceiver designs.
\item \textbf{H-A:} Programmable active (\textbf{A}) SIMs possess the capability to simultaneously reconfigure their amplitude and phase responses through FPGA control. Each meta-atom is integrated with an amplifier, enabling a wide dynamic range for adjusting the amplitude of incident signals. As indicated in Tab. \ref{Hard_survey}, Liu \emph{et al.} \cite{liu2022programmable} fabricated an active SIM operating at $5.4$ GHz. Specifically, five metasurface layers are cascaded with a layer spacing of $10$ cm. Each layer comprises $64$ meta-atoms, which consist of three substrate layers: two $1$ mm thick F4B layers sandwiching a $0.2$ mm thick prepreg layer. Furthermore, one can achieve nonlinear activation functions by operating the meta-atoms in the nonlinear power amplification range. As a result, active SIMs demonstrate the potential to fully execute neural network operations and handle computationally intensive tasks, such as image recognition, directly in the wave domain.
\end{itemize}

\begin{figure*}[!t]
\centering
{\includegraphics[width=18cm]{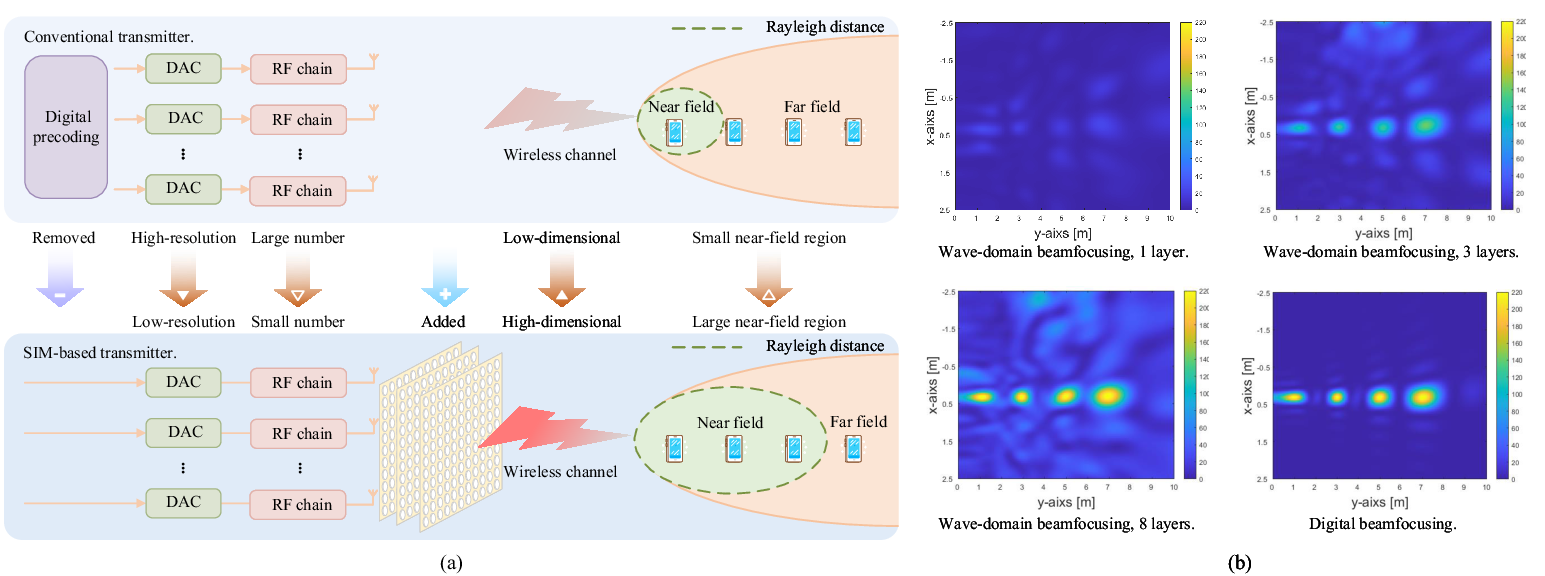}}
\caption{(a) Conventional transceiver vs. SIM-based transceiver. (b) Wave-domain beamfocusing capability of SIM for near-field communications.}
\label{Fig2}
\end{figure*}

In addition, SIMs can be strategically deployed at different locations (\textbf{L}) in communication systems to fulfill diverse tasks. Typically, their thickness is constrained to only a few wavelengths, which enables their seamless integration with practical objects such as the radome of base stations (BS) and windows. Practical deployment scenarios can be categorized into three main types:
\begin{itemize}
\item \textbf{L-T:} When deployed at the transmitter (\textbf{T}), SIMs can directly perform encoding, modulation, and beamforming functions in the wave domain. This significantly enhances signal processing efficiency and reduces computational overhead.
\item \textbf{L-E:} When positioned in the environment (\textbf{E}), SIMs with greater electromagnetic tuning capability can modify the propagation conditions to improve signal strength at target locations, while suppressing interference from undesired directions.
\item \textbf{L-R:} When integrated with the receiver (\textbf{R}), SIMs can efficiently decode incoming signals and extract pivotal information for a variety of sensing and communication applications.
\end{itemize}

\section{Wave-domain Analog Beamforming}
In this section, we elaborate on the potential of SIMs to achieve wave-domain analog beamforming and offer insights into their applications in wireless communications.

\subsection{SIM-assisted Wireless Transceivers}
Typically, conventional MIMO transmitters use digital precoding to mitigate inter-stream interference. However, this requires each transmit antenna to be equipped with a high-resolution digital-to-analog converter (DAC) and a separate RF chain, thus leading to prohibitive hardware costs and complexity, especially for extremely large antenna array (ELAA) systems \cite{zhang2023RHS}. Fortunately, SIM-based transceivers can effectively address these issues. As depicted in Fig. \ref{Fig2}(a), compared to conventional MIMO transmitters, two distinct changes in the SIM-based transceiver architecture are identified:
\begin{itemize}[label={}, labelsep=0em, leftmargin=2.5em]
\item[$(\bm{-})$: ] The high-cost digital baseband precoding is removed.
\item[$(\bm{+})$: ] The low-cost SIM is integrated with the antenna array for achieving wave-domain analog precoding.
\end{itemize}

Accordingly, the transceiver architecture and signal transmission design are affected:
\begin{enumerate}
\item \textbf{Transceiver}: By leveraging SIM to construct multiple parallel interference-free subchannels in physical space, the hardware architecture of MIMO transceivers is significantly simplified. Specifically, when considering practical low-order modulation schemes, each individual data stream can be transmitted from its corresponding transmitting antenna using a low-resolution DAC \cite{an2024magazine}. This scheme also minimizes the number of active RF chains required, enabling multi-stream transmission with significantly reduced hardware expenses \cite{liu2024drl}.
\item \textbf{Transmission}: Compared to conventional antenna arrays, metasurfaces can accommodate a large number of low-cost meta-atoms, thus forming a high-dimensional wireless channel in SIM-assisted systems \cite{yao2024channel}. According to the Rayleigh boundary, the large aperture of metasurfaces would considerably extend the near-field region \cite{papazafeiropoulos2024near}. Consequently, a greater number of users fall within the radiating near-field region (as illustrated in Fig. \ref{Fig2}(a)), allowing for more accurate and efficient beamfocusing to serve multiple users located in the same angular direction.
\end{enumerate}

\begin{figure*}[ht]
\centering
{\includegraphics[width=16.5cm]{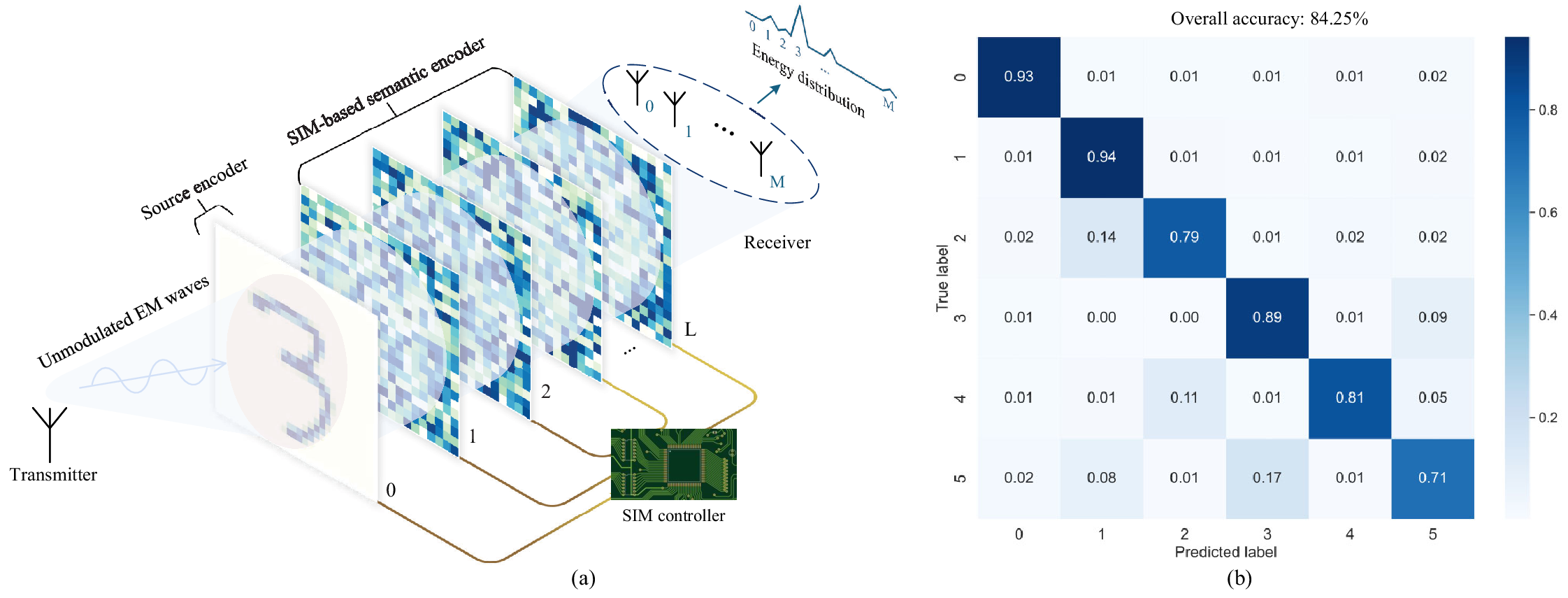}}
\caption{(a) An SIM-assisted semantic communication system, where the SIM transforms electromagnetic waves to a unique beam towards the receiving antenna corresponding to its image class. (b) Confusion matrix of recognizing $6$ digits in the MNIST dataset.}
\label{fig:semantic}
\end{figure*}

\subsection{Relevant Advances and Case Studies}
Again, wave-domain beamforming offers a promising solution for replacing digital precoding by directly manipulating electromagnetic waves, thus leading to higher efficiency and lower costs \cite{an2024magazine}. However, to achieve the expected wave-domain computation functionalities, appropriate configuration of the SIM transmission coefficients is required. To this end, the authors of \cite{liu2024drl} and \cite{papazafeiropoulos2024near} jointly optimized the phase shift patterns of multiple metasurfaces and transmit power allocation for SIM-assisted far-field and near-field multi-user communications, respectively. Specifically, deep reinforcement learning (DRL) and gradient-based optimization methods were utilized to update the transmission coefficients of meta-atoms in SIM. Moreover, Lin \emph{et al.} \cite{lin2024stacked} explored the use of SIMs in multi-beam satellite communication systems, achieving significant reductions in processing delay and computational overhead. In addition, Hu \emph{et al.} \cite{Hu_cellfree} integrated SIMs into cell-free networks to reduce the power consumption of access points without compromising the system sum rate.

To verify the beamfocusing capability of SIM, we consider a downlink communication scenario where an SIM is integrated with the radome of the BS operating at $3$ GHz. Each metasurface layer consists of $225$ meta-atoms, and the layer spacing is $0.01$ m. Four users are placed in the boresight direction of the SIM, spaced $1.5$ m apart. As shown in Fig. \ref{Fig2}(b), the wave-domain beamfocusing capability of the SIM is progressively improved as the number of layers increases. Specifically, a single-layer metasurface fails to focus the desired signals onto the users because of its limited ability to manipulate the incident waves. The SIM having three layers shows improved focusing performance but still results in substantial power leakage into unwanted regions. As the number of metasurface layers increases to $8$, the SIM achieves excellent focusing performance comparable to the conventional digital zero-forcing (ZF) precoding. Benefiting from the enhanced interference cancellation capability of the multi-layer structure, SIMs can effectively generate multiple beams toward the four users' locations.

\section{Semantic Encoder}
In this section, we explore the application of SIMs as semantic encoders. We first elaborate on the underlying operating mechanism and then discuss related research progress.

\subsection{SIM-aided Semantic Encoder}
Semantic communication aims to efficiently transmit task-relevant information and recover the meaning of data with minimum semantic error \cite{huang2024stacked}. In this context, SIM can be utilized for implementing semantic encoding in the wave domain. Unlike conventional semantic communication systems that sequentially process and encode source data, SIM can perform source and semantic encoding automatically as electromagnetic waves propagate through their multi-layer architecture.

To elaborate, Fig. \ref{fig:semantic}(a) shows an SIM-aided semantic communication system, where the input layer is utilized to encode input data, such as images, into the transmission coefficient pattern, while subsequent layers form an electromagnetic neural network to extract task-relevant semantics by processing electromagnetic waves carrying the source information. For an image classification task, the electromagnetic signals are transformed into a unique beam corresponding to the specific image category. As a result, the image recognition task can be readily completed at the receiver by probing the energy distribution across the antenna array with low computational complexity. In a nutshell, this new semantic communication paradigm relying on SIM can significantly reduce latency and energy consumption while maintaining high computational efficiency.

\subsection{Relevant Advances and Case Studies}
In \cite{huang2024stacked}, Huang \emph{et al.} investigated the potential of SIM to encode images and transform them into distinct beam patterns corresponding to the specific image classes. By doing so, the peak-to-average power ratio of the transmit signals was significantly reduced. Additionally, Li \emph{et al.} \cite{li2021spectrally} developed a semantic encoder using a static SIM to convert object information into spectral power distributions. They showed that the designed $3$-layer SIM achieves $96\%$ accuracy in MNIST classification by employing only a single detector.

Next, we study an image classification task to verify the capability of SIM. Specifically, we consider a $7$-layer SIM operating at $28$ GHz, with $441$ meta-atoms per layer to process $6$ categories of handwritten digits in the MNIST dataset. The transmit power and average noise power are set to $40$ dBm and $-104$ dBm, respectively. Other parameters are set in accordance with \cite{huang2024stacked}. Fig. \ref{fig:semantic}(b) shows the confusion matrix for recognizing six handwritten digits, where the SIM is trained using the mini-batch gradient descent algorithm in \cite{huang2024stacked}. It is noted that the SIM with a multi-layer structure can precisely transform information-bearing signals onto their corresponding beams. Specifically, the system achieves an overall accuracy of $84.52\%$. 
By further incorporating nonlinear components, such as feedback-controlled power amplifiers through feedback signals and leveraging their nonlinear range, the performance of SIMs can be further enhanced.

\begin{figure}[!t]
\centering
\subfloat[]{\includegraphics[width=7.5cm]{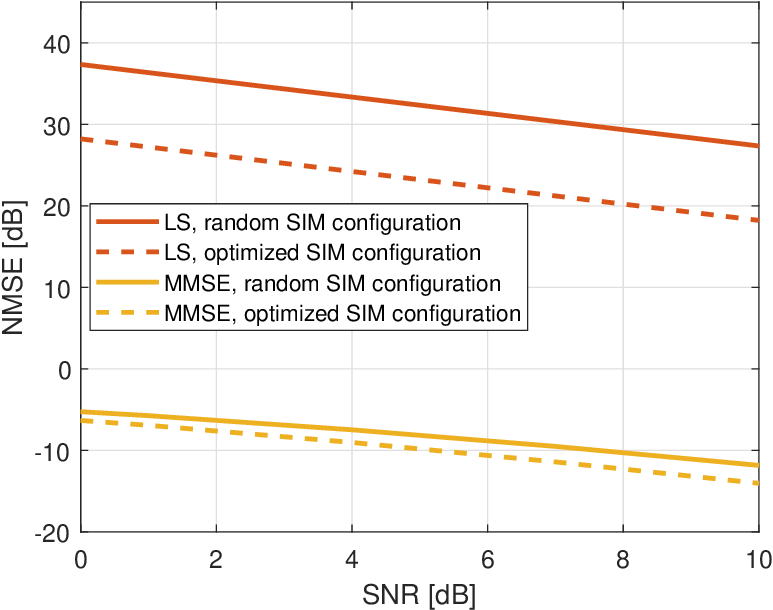}}\\
\subfloat[]{\includegraphics[width=7.5cm]{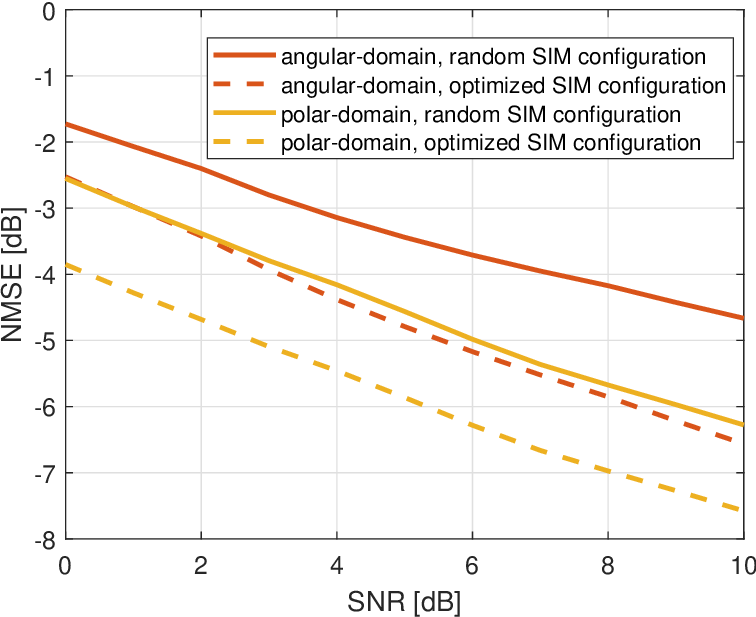}}
\caption{NMSE versus SNR by employing (a) conventional channel estimation schemes, and (b) SBL-based channel estimation schemes, respectively.}
\label{fig:channel_estimation}
\end{figure}

\section{Channel Estimation}
Accurate channel estimation is critical to achieve various desired wave-domain computation functionalities. In this section, we discuss the unique challenges and potential solutions in channel estimation for SIM-assisted communication systems.

\subsection{New Features and Challenges}
To efficiently perform the desired communication tasks, typically, a large number of meta-atoms are deployed on each metasurface of an SIM, thus resulting in high-dimensional channels. However, practical SIM-aided systems typically have a limited number of RF chains at the BS, leading to an underdetermined channel estimation problem and exorbitant pilot overhead. Additionally, due to the significantly larger aperture of SIM structures, conventional channel estimation methods based on planar wavefront are no longer applicable in the radiating near field. Compared to the conventional single-layer system, the multi-layer SIM provides additional optimization degrees of freedom (DoF) for enhancing beamforming flexibility and precision. Remarkably, the estimation accuracy can be enhanced through adaptive adjustment of SIM phase parameters. Furthermore, sparse channel estimation techniques in the polar domain present a viable approach to reconcile SIM's improved beamforming performance with limited pilot overhead, while the development of adaptive estimation protocols that dynamically adjust the trade-off between estimation accuracy and pilot signaling overhead requires further investigation.

\subsection{Relevant Advances and Case Studies}
To address the underdetermined channel estimation problem, Yao \emph{et al.} \cite{yao2024channel} developed an estimation protocol that leverages multiple observations of the uplink pilot signals to estimate the wireless channels. In addition to applying the conventional least squares (LS) and minimum mean square error (MMSE) methods, they proposed a reduced-subspace LS estimator, which leverages the rank deficiency of spatial correlation matrices to filter the channel estimates, thereby improving the estimation accuracy.
Moreover, hybrid digital and wave-domain \cite{nadeem2024channel_estimate} were proposed to optimize SIM phase shifts to minimize the mean square error (MSE) of the channel estimator. It has been demonstrated that the hybrid approach with fewer RF chains achieves comparable estimation performance to fully digital systems. In addition, the impacts of electromagnetic coupling effects in channel modeling and estimation were systematically analyzed in \cite{nerini2024physically}.

Next, we investigate the channel estimation of SIM-assisted systems. Specifically, a $6$-layer SIM with $100$ meta-atoms per layer is considered. The system operates at $28$ GHz, and other system parameters remain the same with \cite{yao2024channel}. Fig. \ref{fig:channel_estimation}(a) evaluates the normalized MSE (NMSE) of the LS and MMSE estimation techniques under different signal-to-noise ratio (SNR) conditions, where the transmission coefficients of SIM are optimized according to the method in \cite{yao2024channel}. Note that after optimizing the SIM phase shift configuration, both conventional LS and MMSE estimators demonstrate significant performance improvements, with the LS estimator exhibiting a nearly $10$ dB performance gain due to the capability of the SIM in achieving favorable channel tuning.

Fig. \ref{fig:channel_estimation}(b) evaluates the NMSE performance of sparse Bayesian learning (SBL)-based channel estimation techniques for SIM-assisted near-filed communication systems, where we consider a $4$-layer SIM operating at $30$ GHz. Each layer contains $1,600$ meta-atoms with half-wavelength spacing at the receiver. Since the near-field channels exhibit severe energy dispersion in the angular domain, leveraging the polar domain transformation would enhance the sparsity and enable the identification of dominant propagation paths’ support sets and complex gains. Moreover, the SIM response can also be optimized to reduce the coherence between columns in the dictionary matrix, yielding an average NMSE reduction of $1$ dB compared to that without optimizing SIM. These results highlight that the multi-layer SIM architecture provides extra tuning DoF to improve the estimation accuracy.

To provide a systematic overview, we summarize recent SIM-related research advances in Tab. \ref{Summary_SIM_Table}.
\begin{table*}[!t]
\centering
\renewcommand{\arraystretch}{1.5}
\caption{A survey of recent advances in SIM-assisted wireless communications.}
\label{Summary_SIM_Table}
\begin{tabular}{l|l|l|l|l|l|l|l}
\hline
Ref. & \makecell[l]{SIM's function} & Scenario/Task & \makecell[l]{SIM's\\ location} & \makecell[l]{Hardware} & \makecell[l]{Optimization variables} & \makecell[l]{Objective function} & Optimization method \\\hline
\cite{Hu_cellfree} & \multirow{7}{*}{\makecell[l]{Wave-domain\\beamforming}} & Cell-free network & L-T & H-P & \rule{0pt}{3.5ex}\makecell[l]{$\ominus$ SIM phase shift $\mathcal{C}$\\$\diamondsuit$ Power allocation $\mathcal{C}$} & Sum-rate & \makecell[l]{AO} \\ \cline{1-1} \cline{3-8} 
\cite{liu2024drl} & & Multiuser MISO & L-T & H-P & \rule{0pt}{4.5ex}\makecell[l]{$\ominus$ SIM phase shift $\mathcal{C}$\\$\diamondsuit$ Power allocation $\mathcal{C}$} & Sum-rate & DRL \\ \cline{1-1} \cline{3-8} 
\cite{papazafeiropoulos2024near} & & \makecell[l]{Near-field network} & L-T & H-P & \rule{0pt}{3.5ex}\makecell[l]{$\ominus$ SIM phase shift $\mathcal{C}$\\$\diamondsuit$ Power allocation $\mathcal{C}$} & \makecell[l]{Weighted sum-rate} & Block coordinate descent \\ \cline{1-1} \cline{3-8} 
\cite{lin2024stacked} & & Satellite network & L-T & H-P & \makecell[l]{$\ominus$ SIM phase shift $\mathcal{C}$ \\ $\diamondsuit$ Power allocation $\mathcal{C}$} \rule{0pt}{3.5ex} & Sum-rate & \makecell[l]{AO} \\ \cline{1-1} \cline{3-8} 
\cite{hassan2024_efficient} & & SISO & L-E & H-P & $\ominus$ SIM phase shift $\mathcal{C}$, $\mathcal{D}$ & \makecell[l]{Received power } & Gradient descent \\ \hline
\cite{li2021spectrally} & \rule{0pt}{2ex}\multirow{4}{*}{\makecell[l]{Semantic\\encoder}} &Image classification & L-T & H-S & \rule{0pt}{3.5ex}\makecell[l]{$\ominus$ SIM phase shift $\mathcal{D}$\\$\oslash$ SIM amplitude $\mathcal{D}$} & Cross entropy & Gradient descent \\ \cline{1-1} \cline{3-8}
\cite{liu2022programmable} & & \makecell[l]{Image classification } & L-T & H-A & \rule{0pt}{3.5ex}\makecell[l]{$\ominus$ SIM phase shift $\mathcal{D}$\\$\oslash$ SIM amplitude $\mathcal{D}$} & Classification accuracy & Gradient descent \\ \cline{1-1} \cline{3-8} 
\cite{huang2024stacked} & & \makecell[l]{Image classification } & L-T & H-P & \rule{0pt}{3.5ex}\makecell[l]{$\ominus$ SIM phase shift $\mathcal{C}$\\$\oslash$ SIM amplitude $\mathcal{C}$} & Cross entropy & Gradient descent \\ \hline
\cite{yao2024channel} & \rule{0pt}{2ex}\multirow{2.5}{*}{\makecell[l]{Channel\\estimation}} & Multiuser MISO & L-R & H-P & $\ominus$ SIM phase shift $\mathcal{C}$ & MSE & Codebook \\ \cline{1-1} \cline{3-8}
\cite{nadeem2024channel_estimate} & & Multiuser MISO & L-R & H-P & \rule{0pt}{3.5ex}\makecell[l]{$\ominus$ SIM phase shift $\mathcal{C}$} & NMSE & Gradient descent \\ \hline
\cite{wang2024multi} & \multirow{2}{*}{ISAC} & Multiuser MISO & L-E & H-P & \rule{0pt}{4.5ex}\makecell[l]{$\ominus$ SIM phase shift $\mathcal{C}$\\$\Box$ Transmit beamforming $\mathcal{C}$} & \makecell[l]{CRB} & AO and SDR \\ \cline{1-1} \cline{3-8} 
\cite{TVT_2024_Li_Transmit} & & Multiuser MISO & L-T & H-P & \rule{0pt}{3.5ex}\makecell[l]{$\ominus$ SIM phase shift $\mathcal{C}$}& \makecell[l]{Sum-rate, MSE} & Gradient descent \\ \hline
\multicolumn{8}{l}{{ MISO: Multiple-input single-output. SISO: Sinple-input single-output. SDR: Semidefinite relaxation. CRB: Cramér-Rao Bound. }} \\
\multicolumn{8}{l}{{ $\mathcal{C}$: Continuous variables. $\mathcal{D}$: Discrete variables.}} \\
\hline
\end{tabular}
\end{table*}

\section{Research Opportunities}
\begin{figure*}[!t]
\centering
{\includegraphics[width=18cm]{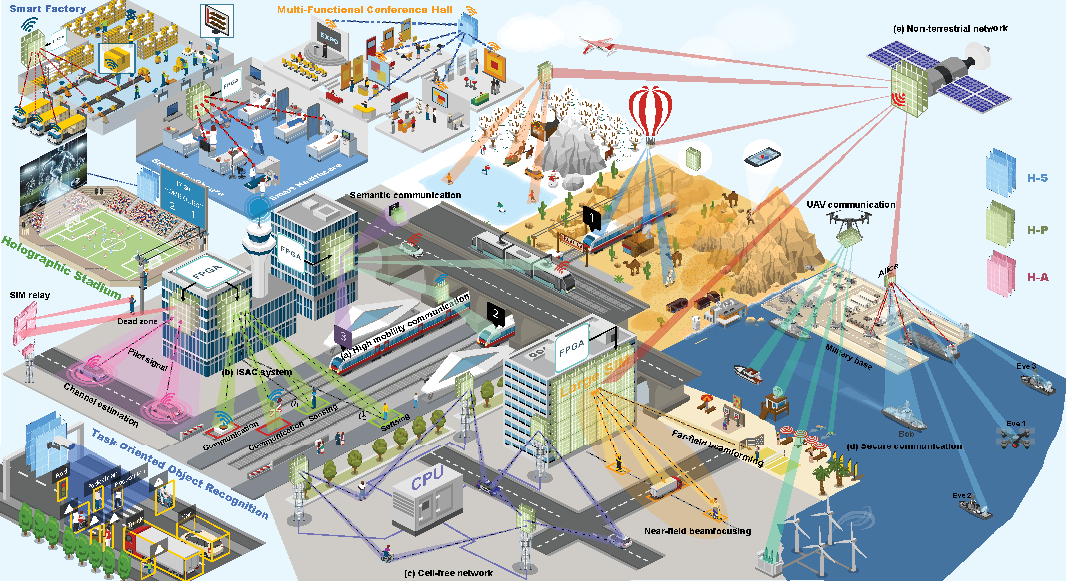}}
\caption {A grand vision of SIM applications.}
\label{SIM_Scenario}
\end{figure*}

In this section, we elaborate on future research opportunities for integrating SIM into wireless networks and discuss its synergies with emerging wireless technologies. Specifically, Fig. \ref{SIM_Scenario} presents a grand vision for promising applications of SIM, where several promising research directions are illustrated.
\begin{itemize} 
\item [(a)] \textbf{SIM-assisted High Mobility Communication:} The ultra-fast signal processing capability of SIM makes it particularly effective for high-speed mobile communications. For instance, SIM can implement wave-domain equalization for effectively mitigating interference between subcarriers caused by Doppler shifts. When deployed in bridge tunnels, SIMs can act as relays to enhance signal quality and ensure reliable communication links, even in high-mobility environments. In addition, as high-speed trains enter or exit stations, SIM can be utilized to identify train numbers and control the track switches, ensuring smooth and efficient operations. When implemented in fixed-wing platforms, SIM can achieve simultaneous real-time beam tracking and adaptive interference mitigation, thus improving outage performance under high-velocity flight regimes.
\item [(b)] \textbf{SIM-assisted Integrated Sensing and Communication:} SIM can generate dual-function waveforms to achieve simultaneous communication and sensing with reduced hardware costs \cite{TVT_2024_Li_Transmit}. As illustrated in Fig. \ref{SIM_Scenario}, SIM can be utilized to detect track intrusions near platforms, identify individuals crossing the tracks, and provide real-time alerts to train operators. Meanwhile, SIM can receive and process wireless data from train-tracking sensors, allowing them to adjust signal lights and optimize train flow to improve the efficiency of station services.
\item [(c)] \textbf{SIM-assisted Cell-free Network:} By offloading complex calculations at BSs, SIM can simplify transceiver architectures and reduce extra hardware costs \cite{Hu_cellfree}. This makes it more attractive to deploy a large number of access points and small BSs in wireless networks. As shown in Fig. \ref{SIM_Scenario}, in an SIM-assisted cell-free network, a central processing unit (CPU) manages dynamic resource allocation and analog beamforming to eliminate blind spots and handovers, even in areas where users are far from access points, thus ensuring a seamless communication experience.
\item [(d)] \textbf{SIM-assisted Physical Layer Security:} By leveraging wave-domain signal processing, SIM can achieve precise beamforming to concentrate the desired signal on the intended users. This capability is particularly effective in enhancing physical layer security. For instance, in maritime scenarios shown in Fig. \ref{SIM_Scenario}, an SIM ensures that authorized vessels (Bobs) receive a strong signal, while potential eavesdroppers (Eves) in the surrounding sea regions are largely prevented from intercepting the signal, thereby facilitating secure communication for maritime applications.
\item [(e)] \textbf{SIM-assisted Non-terrestrial Network:} SIM (H-P) lends itself to efficient deployment in non-terrestrial networks (NTN), thus ensuring low-maintenance operation and a long service life \cite{lin2024stacked}. Compared to traditional satellite payloads, SIM enables task-switching, e.g., wide-area coverage or directional beam, through programming without requiring hardware reconfiguration. As shown in Fig. \ref{SIM_Scenario}, integrating SIM into vertical heterogeneous networks (vHetNet) enables efficient communication across multiple altitude layers, including satellites, high-altitude platform stations, and various aerial platforms. This multi-layer architecture can effectively extend network coverage to remote areas such as deserts and snow-covered regions. Thus, the deployment of SIM in NTN can provide heterogeneous communication links in emergency scenarios or during scientific expeditions, collectively establishing reliable messaging and data exchange capabilities in harsh environments.
\end{itemize}

Moreover, SIMs can be integrated into various cellular and Wi-Fi networks to support massive access via wave-domain beamforming and unlock a range of exciting applications, such as smart factories, multi-functional conference halls, holographic stadiums, smart healthcare, and intelligent transportation systems, ultimately paving the way for the future wireless networks.

\section{Implementation Challenges of SIMs}
Despite the great promise of SIM, several challenges need to be overcome for its practical implementation. In this section, we discuss four key challenges of deploying SIMs in wireless networks and offer insights into potential solutions.
\subsection{Practical Hardware Imperfections}
\textit{\textbf{Challenges}}: The performance of SIM depends heavily on the accurate models. However, in practice, various factors can lead to hardware imperfections. These include voltage jitter in control circuits \cite{liu2022programmable}, meta-atom inconsistencies caused by manufacturing or assembly flaws, as well as adverse coupling between metasurfaces and meta-atoms. Additionally, the actual propagation coefficients between layers may not align with the values predicted by Rayleigh-Sommerfeld diffraction theory in numerical models \cite{an2024magazine}. As a result, these hardware imperfections compromise the capability of SIMs to perform wave-domain signal processing tasks.

\textit{\textbf{Solutions}}: The calibration of inter-layer propagation coefficients can be achieved by analyzing the difference between the actual output of SIMs and the target output and designing effective compensation techniques \cite{liu2022programmable}. Additionally, addressing coupling effects requires the development of physically consistent models. In order to enhance the robustness of wave-domain computations, it is suggested to incorporate datasets that reflect the variability of hardware imperfections during the training phase. This method has the potential to substantially improve the reliability and resilience of SIMs under practical conditions \cite{li2021spectrally}.

\subsection{Efficient SIM Configuration}
\textit{\textbf{Challenges}}: In SIM-assisted communication systems, a major challenge is designing efficient SIM configuration algorithms that can adapt to dynamic environments and varying task requirements. Achieving optimal performance typically requires the joint optimization of SIM phase shifts and resource allocation strategies. However, this often leads to a non-convex problem characterized by multiple highly coupled variables. Currently, a common approach is to use alternating optimization (AO) algorithms, which, however, may suffer from high computational overhead, slow convergence, and sensitivity to initial parameter settings \cite{lin2024stacked}. 

\textit{\textbf{Solutions}}: In order to tackle the complex non-convex discrete optimization problem, a straightforward approach is to relax some constraints and transform the original problem into a more tractable form. The solution is then projected onto a feasible domain that satisfies the practical hardware limitations. Additionally, to reduce the computational load of optimizing a large number of transmission coefficients, advanced machine learning techniques such as DRL can be employed, while the specific algorithm design requires further investigation.

\subsection{Performance Analysis}
\textit{\textbf{Challenges}}: A solid theoretical framework is essential for assessing the performance limits of SIMs at both device and network levels. In point-to-point communication scenarios, such an analytical framework enables the evaluation of trade-offs among computational capabilities, energy consumption, and system complexity. However, the challenges become more pronounced in large-scale SIM deployments, where network-wide interference management and scalability issues introduce significant complexity. These interrelated factors, coupled with the high-dimensional parameter space of cascaded metasurfaces, complicate the derivation of exact, closed-form solutions, especially in dynamic wireless channels with continuously varying task requirements and heterogeneous service demands.

\textit{\textbf{Solutions}}: Since the response of SIMs relies on cascaded matrix operations involving wave propagation and transmission coefficients, matrix decomposition theory may become a key tool to characterize its computational capabilities. For large-scale deployments, it is crucial to develop distributed optimization frameworks that can efficiently manage network-level interference while maintaining scalability. By extracting key semantic information across different network nodes, context-aware resource allocation may become feasible to adapt to varying channel conditions and application requirements.

\subsection{SIM Deployment}
\textit{\textbf{Challenges}}:
In practical systems, different application scenarios impose diverse requirements on the power consumption, response speed, and processing capabilities of SIMs. Therefore, identifying the optimal placement, appropriate hardware types, and suitable physical dimensions of SIMs to fulfill specific service requirements constitutes a fundamental challenge.

\textit{\textbf{Solution}}: In areas with stable conditions and simple functional requirements, SIMs (H-S) can be deployed. However, in more complex communication environments, e.g., densely populated urban areas, deploying large-aperture SIMs combined with optimized beam scheduling and control enables precise manipulation of electromagnetic waves to mitigate the effects of small-scale fading.

\section{Conclusion}
The integration of SIMs into wireless communication networks demonstrated immense promise, thanks to its ability to perform computations in the wave domain. This article presented a comprehensive overview of SIM technology, covering its fundamental principles, hardware architectures, and prospective applications. We also identified emerging research opportunities for applying SIMs in various scenarios. In addition, we analyzed the significant challenges that come with implementing SIMs and provided insights into future research directions for realizing low-latency and energy-efficient computation.

\bibliographystyle{IEEEtran}
\bibliography{Ref}

\newpage
\section*{Biographies}

\noindent\textbf{Hao Liu} [S] is currently pursuing the Ph.D. degree with the School of Information and Communication Engineering at the University of Electronic Science and Technology of China (UESTC), Chengdu, China.
\vspace{1em}

\noindent\textbf{Jiancheng An} [M] is currently a Research Fellow with the School of Electrical and Electronics Engineering, Nanyang Technological University, Singapore.
\vspace{1em}

\noindent\textbf{Xing Jia} [S] is currently pursuing the Ph.D. degree with the School of Information and Communication Engineering at the UESTC, Chengdu, China.
\vspace{1em}

\noindent\textbf{Lu Gan} [M] is currently a Professor with the School of Information and Communication Engineering at the UESTC, Chengdu, China.
\vspace{1em}

\noindent\textbf{George K. Karagiannidis} [F] is a Professor and the Head of Wireless Communications and Information Processing Group in the Department of Electrical and Computer Engineering, Aristotle University of Thessaloniki, Thessaloniki, Greece, and a Faculty Fellow with the Artificial Intelligence \& Cyber Systems Research Center, Lebanese American University, Beirut, Lebanon.
\vspace{1em}

\noindent\textbf{Bruno Clerckx} [F] is currently a Professor and the Head of the Wireless Communications and Signal Processing Laboratory and the Head of the Communications and Signal Processing Group, Department of Electrical and Electronic Engineering, Imperial College London, U.K.
\vspace{1em}

\noindent\textbf{Mehdi Bennis} [F] is currently a tenured Full Professor with the Centre for Wireless Communications, University of Oulu, Finland, and the Head of the Intelligent COnnectivity and Networks/Systems Group (ICON). His research interests include radio resource management, game theory, and distributed AI in 5G/6G networks. 
\vspace{1em}

\noindent\textbf{M\'{e}rouane Debbah} [F] is a Professor at Khalifa University of Science and Technology in Abu Dhabi and founding Director of the KU 6G Research Center. He is known for his work on Large Language Models, distributed AI systems for networks, and semantic communications.
\vspace{1em}

\noindent\textbf{Tie Jun Cui} [F] is the director of the State Key Laboratory of Millimeter Waves, and the Founding Director of the Institute of Electromagnetic Space, Southeast University. He is an academician of the Chinese Academy of Science. His research interests include metamaterials and computational electromagnetics.
\vspace{1em}

\end{document}